# Evaluating Tenant-Landlord Tensions Using Generative AI on Online Tenant Forums

Xin Chen[1], Cheng Ren[2], Timothy A Thomas[3]

**Abstract**

Tenant-landlord relationships exhibit a power asymmetry where landlords' power to evict the tenants at a low-cost results in their dominating status in such relationships. Tenant concerns are thus often unspoken, unresolved, or ignored and this could lead to blatant conflicts as suppressed tenant concerns accumulate. Modern machine learning methods and Large Language Models (LLM) have demonstrated immense abilities to perform language tasks. In this study, we incorporate Latent Dirichlet Allocation (LDA) with GPT-4 to classify Reddit post data scraped from the subreddit r/Tenant, aiming to unveil trends in tenant concerns while exploring the adoption of LLMs and machine learning methods in social science research. We find that tenant concerns in topics like fee dispute and utility issues are consistently dominant in all four states analyzed while each state has other common tenant concerns special to itself. Moreover, we discover temporal trends in tenant concerns that provide important implications regarding the impact of the pandemic and the Eviction Moratorium.

**Introduction**

Tension in tenant-landlord relationships emerges frequently due to the intrinsic power dynamics within this type of relationship. Specifically, there is a power asymmetry in tenant-landlord relationships due to the many powers that the landlord has over the tenant, especially the power to evict the tenant at a low filing fee (Gomory et al., 2023). Tenant-landlord tension may arise from many circumstances, including property maintenance, rent setting, leasing, and contracting (Kapolka, 2022). Tenant concerns in those circumstances can greatly impair their rental and living experience. However, the relatively disempowered status of tenants in tenant-landlord relationships causes many of their concerns to remain unspoken, unresolved, or ignored by the landlord (Chisholm et al., 2020). When the tension between tenants and landlords escalates into blatant conflicts, this may result in the eviction of the tenant by the landlord, which brings various negative repercussions to the tenant's mental health, physical health, career, and children's development (Kapolka, 2022). Thus, it is of vital importance to unravel the main risks of tensions

1. Stanford University; Stanford, CA
2. University at Albany, State University of New York, Albany, NY
3. University of California, Berkeley; Berkeley, CA

Corresponding author:

Cheng Ren, 135 Western Avenue, Albany, New York, USA. Email: cren@albany.edu

in tenant-landlord relationships and understand dominant tenant concerns. A thorough and up-to-date understanding of this would be invaluable to the creation and reform of housing policies and housing aid programs to moderate the power imbalance in tenant-landlord relationships.

The development of modern Large language models, which demonstrate enormous zero-shot abilities in language tasks, brings in new possibilities in tenant-landlord relationship research (Si et al., 2023). Zero-shot classification, in particular, refers to the classification task where possible values for the outcome variable include values that were omitted from the training set (Palatucci, 2009). Ziems et al. (2023) evaluated various LLMs on common language tasks in computational social sciences, identifying their potential power to transform the field. In fact, many researchers in social sciences have already adopted LLMs to facilitate their research, taking advantage of LLM's capabilities to facilitate labeling, classification, and simulation (Chiu and Alexander, 2021; Egami et al., 2023; Guo, 2023; Paoli, 2023). Nevertheless, there remain very few or no studies that adopt modern LLMs to facilitate research in tenant-landlord relationships. In this study, we seek to utilize the power of LLMs, specifically GPT-4, in combination with traditional machine learning methods to advance research in tenant-landlord relationships using a novel dataset scraped from Reddit.

The main research question addressed in the study is: How do tenant concerns vary geographically and temporally? The aims of the study are two-fold. The first aspect is focused on the method where (1) we would like to explore ways to efficiently incorporate machine learning methods and modern large language models to facilitate tenant-landlord research. The second aspect of this study's aims concentrates on advancing knowledge in tenant-landlord research, including (2) extracting geographic homogeneity and heterogeneity in tenant concerns and (3) exploring temporal trends in tenant concerns to evaluate the influence of exogenous events and policies like the pandemic.

**Literature Review**

*Tenant-Landlord Relationships*

This study analyzes tenant concerns and tenant-landlord relationships through topic modeling and zero-shot classification on Reddit posts, aiming to innovate upon the traditional methods adopted by researchers in housing and tenant-landlord relationships. One mainstream methodology in tenant research is surveys and interviews. For instance, the American Housing Survey (AHS) from the US Census Bureau publishes housing-related statistics biennially from survey results (Iriondo, 2022). The AHS studies a broad scope of topics in housing, including tenant situations but also examines the status of house owners. Another methodology in tenant research utilizes mathematical models. Coulson et al. (2020) approached the study of tenant-landlord relationships from a mathematical modeling perspective, where they modeled the tenant-landlord relationship using search theory to predict eviction rates due to changes in rent. Approaching the modeling of tenant situations with a more empirical emphasis, the Urban Displacement Project built models of housing risk indices to provide a geographic profile of risks faced by tenants (Chapple et al., 2021). However, the first method is costly in terms of both time and money if a large sample is needed, and the second method did not explore the rich textual information regarding the tenant experience.

In the meantime, evictions remain an urgent topic in Tenant-Landlord relationships. Researchers study the nature and causes of evictions and propose social or legal reforms to alleviate the issue of tenant evictions. The majority of evictions take place due to rent arrears, conflicts between landlords and tenants, violation of property rules by tenants, or even policy influence (Holl et al., 2016; Desmond, 2016; Griswold et al., 2023; Thomas et al., 2024). Tenant-landlord relationships could trigger many of the reasons for eviction mentioned above. Moreover, the impact of legal aids during the eviction process also plays a role in eviction research. Seron et al. (2001) observed from randomized experiment in New York's Housing court that legal counseling and representation by volunteer attorneys resulted in a lower number of eviction warrants. This case also indicates that the fault could be on either the tenant's or the landlord's side, which is worth exploring in detail of the relationship before.

As mentioned above, understanding Tenants' concerns is significant in tenant research as many concerns are correlated with evictions and a better view of the concerns can help propose appropriate policies to enhance tenant experiences and tenant-landlord relationships. Studying tenant concerns is made more complex by the heterogeneity among different states in housing and eviction policies. Merritt and Farnworth (2021) claimed that a state with a more "tenant-friendly policy environment" is correlated with lower eviction rates. Tenant experiences also vary in different time periods due to their close relationship with the overall economic well-being of the nation. During a global pandemic like COVID-19, studies showed that in many regions, there is a negatively impacted rental experience related to financial insecurity (Bates, 2020). Additionally, landlord power erodes policy protection of tenants during the pandemic (Kapolka, 2022). Structural inequality also becomes amplified and reinforced during economic recessions (Byrne and Sassi, 2022). Thus, the study of landlord-tenant relationships should consider time-related events and pre-existing geographic features that both homogeneously and heterogeneously affect tenants' rental experiences.

*Large Language Models in Social Science Research*

The use of traditional machine learning methods is constrained in many aspects (Ziems et al., 2023). In supervised methods, a large amount of training data is required to train a model with satisfactory performance. On the other hand, unsupervised methods, which do not involve training data, lack interpretability. LLMs can help reconcile the drawbacks of both supervised and unsupervised methods, by generating classifications, summaries, responses, and explanations comparable to human-level accuracy while requiring zero training data due to their zero-shot capabilities (Bang et al., 2023).

Large Language Models, or LLMs, are pre-trained on unlabeled text data and used for various downstream tasks (Si et al. 2023). As language models scale up in aspects like the number of parameters or amount of training data, their performances in downstream tasks are enhanced, which we call emergent abilities Wei et al. (2022). Common LLMs nowadays include Google's BERT and OpenAI's GPT-3.5 and GPT-4. A large part of this study relies on OpenAI's GPT-3.5 and GPT-4 API.

Many researchers in computational social sciences have already taken advantage of LLMs to expand the scope and scale of their research. LLMs can be used to classify texts zero-shot and Chiu and Alexander (2021) used GPT-3.5 to classify racist and sexist texts and concluded that LLMs can, after further development, eventually play a role in countering hate speech. Egami et al. (2023) similarly used the classification ability of LLMs to label documents, which were further used for downstream statistical inference in social science. Guo (2023) took advantage of GPT-3.5's generative ability to simulate strategic games in economics, exhibiting its potential power to advance research in game theory. Paoli (2023) experimented on the emulative ability of GPT-3.5 to conduct the commonly adopted semantic analysis approach in social sciences, suggesting LLMs' ability to aid the qualitative research process in social sciences.

*Reddit in Social Science Research*

The use of Reddit and other social media platforms as the source of data has gained attention from modern researchers. Reddit has become a rich source of data for the use of researchers (Amaya et al., 2021). Jamnik and Lane (2017) discovered that volunteer responses from Reddit can be used as high-quality datasets for social science research. Wed scraping technology enabled the convenient use of Reddit posts as an organic source of data for social sciences researchers (Kairam et al, 2024). Hintz and Betts (2022) provided a model script for scraping Reddit posts. The development of natural language processing techniques further enabled researchers to systematically analyze texts from Reddit and uncover insights into their own research topics, making Reddit data more valuable. Choudhury and De (2014) performed a semantic analysis of Reddit posts on mental health to reveal users' tendency to self-disclose on social media. The authors also manipulated metadata of Reddit posts to explain factors of social support on Reddit. Using a specific Reddit group can also effectively constrain the texts on topics of interest for the researchers. For instance, McDowall et al. (2023) used topic modeling to reveal insights into family planning research from the r/birthcontrol Reddit group. Pleasants et al. (2023) similarly analyzed texts from the r/birthcontrol sub-Reddit to explore language use in contraceptives, discovering the value of the group as a space to discuss contraceptive-related topics not addressed in clinical counseling.

**Methods**

The methods include several key components: data collection, data preprocessing, topic exploration, topic classification, and analysis (see Figure 1). In the following subsections, we will go through the details of each key step.

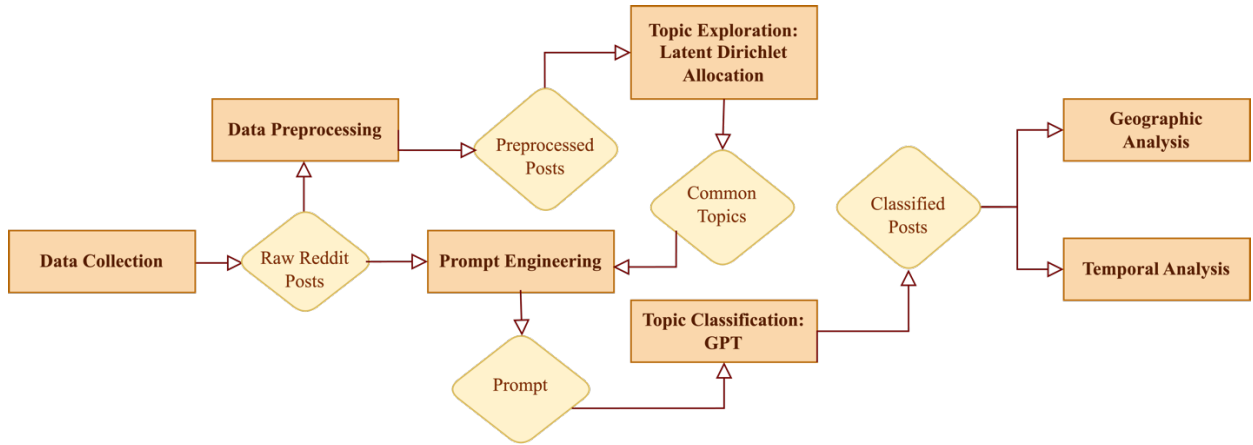

Figure 1. Methodology Flowchart

*Data Collection*

The data was scraped from the Pushshift API (Application Programming Interface) for Reddit (McDowall et al., 2023). PushShift is a data platform for social networking sites that started to collect Reddit data in 2015. A Python package called PSAW was used as an API wrapper to communicate with Pushshift. PushShift Reddit API team noted that this API is faster and easier to use comparing the official Reddit API. Our primary dataset consists of Reddit posts from the subreddit r/Tenant—an online community group designed to discuss landlord/tenant problems. This subreddit is among the top 3% of all communities in size on Reddit, with 61,000 members (12/30/2024). There are a total of 5,896 posts with text collected from 2015/08/06 to 2023/04/30. The data ends on 2023/04/30 because Reddit has rescinded their API access for Pushshift since then[2]. Although they initiated the researcher API, only a select few individuals were chosen for the beta test. Additionally, we also excluded any posts that were either removed by Reddit or moderators (e.g. "[removed]") or deleted by the poster (e.g. "[deleted]"), or without any text. After these steps, the sample size decreased from 5,896 to 4,303. On average, each post receives 4.6 comments.

*Preprocessing*

Prior to delving into the textual information, we ran some essential text preprocessing steps, including eliminating stopwords and applying lemmatization through the NLTK package (e.g. tenants to tenant) for later use in topic modeling. Then the posts were converted into tokens (words). Note that stopwords removal and lemmatization are only applied to the input to the LDA model for topic modeling (described in the next section), and such preprocessing is not done to the input to GPT.

Meanwhile, when people post their experiences, they may indicate their location in the title or the text. Following is an example of how the location could be included in the post: "[US − IN] I've been living in my apartment for a year and a half... ". Thus, we also used name entity recognition to extract geographic locations from each post. We applied the pre-trained model "dslim/bert-large-NER" to extract the location information from the title or body of text, which has an f1-score of 91.7, precision of 91.2, and recall of 92.3 reported in the model documentation (Tjong Kim Sang and De Meulder, 2003). Since the comparison will be conducted at the state level, all location data (e.g., "San Francisco", "US, CA", "SF Bay Area") have been geocoded from text to latitude and longitude using the Azure map server. Subsequently, all geographic location data will be assigned a state name if it falls within the United States, which comprises approximately 38.9% of all texts (N=1,676). This is achieved by merging the data with a shapefile from the US Census. Additionally, around 6.9% (N=296) of the total texts include locations but outside of the US, such as in Ontario, Canada(N=128), England(N=39), British Columbia, Canada(N=29).

---

[2] Reddit Data API Update: Changes to Pushshift Access
https://www.reddit.com/r/modnews/comments/134tjpe/reddit_data_api_update_changes_to_pushshift_access/

### *Exploration of Common Topics: Topic Modeling*

After preprocessing the Reddit Posts into tokens, we applied topic modeling to do an exploratory analysis of the main topics discussed in the posts. To model topics in the Reddit posts, we used Latent Dirichlet Allocation (LDA), which is a generative probabilistic model for collections of discrete data such as text corpora (Blei et al., 2003).

To determine the optimal number of topics for the LDA model, we computed the coherence score on the dataset varying the number of topics from 1 to 25 and found that at around 11 topics, the coherence score graph started to plateau and we decided to model 11 topics from the dataset. We then converted the Reddit Posts into a text corpus and input this into the LDA algorithm for a model to be fitted. The resulting model is a three-level hierarchical Bayesian model, in which each item of a collection is modeled as a finite mixture over an underlying set of topics. Each topic is, in turn, modeled as an infinite mixture over an underlying set of topic probabilities (Blei et al., 2003). After fitting the LDA model, we extracted 11 topics from the posts where a distribution of terms constituted each topic. This allowed us to interpret the meaning of each topic in the context of tenant-landlord relationships and tenant concerns. Nevertheless, LDA is less effective for short texts especially informal texts on social media (Hong and Davison, 2010). Thus, we seek more accurate classification using the newly developed large language model: GPT-4.

### *Zero-shot Classification*

In this study, we used the API of GPT-3.5 and GPT-4 from OpenAI to perform zero-shot classification in batches. We chose to use GPT to represent the capabilities of LLMs in this study due to its excellent performance as well as accessibility among the general public compared to other LLMs. In the results, we deploy GPT-4 without an intermediate summary step, which exhibited the best performance when compared to GPT-3.5 and when adding a summary step. This section lays out the procedure adopted to select this model and prompt for the task of post-text topic classification.

On the model selection dimension, the decision to make is on model selection between the two versions of OpenAI's GPT. GPT is trained to take in instructions written in natural language called prompts. Semantically similar prompts may elicit varied performances and results and thus the prompts need to be engineered to yield results in satisfactory quality and standardized format (Perez et al., 2021). There are no systematic procedures that guide prompt engineering. Instead, we carefully designed two types of prompts, one with and one without a summary, based on best common practices and improved them through trial and error.

We then conducted experiments using both models and both types of prompts by running each model over the validation set (described in the next section) using each type of prompt, with three repetitions for each combination which is similar to having three research assistants to make the choices and combining the results. For each combination of model type and prompt type, we computed the average recalls of the three repetitions to compare the efficacy of each type of prompt on each model to find the best combination of model and prompt.

When prompted with a summary, we broke down the task of Reddit post classification into a two-step process. The first step was prompting GPT to summarize each Reddit text within 10 words. This pass of text processing is aimed at effectively removing noise from verbose texts and enhancing classification performance (Tang et al, 2015; Ma et al, 2018; Deng et al., 2022). The prompt used in this stage is as follows, where we substituted the original Reddit post into the square bracket position for each post:

'*Summarize the following text in 10 words: [original post].*'

10 of the 15 topics in the prompt were obtained from interpreting the LDA results. The topic "Legal Advice" extracted by LDA is not included as most posts are posted for seeking advice related to rental. The rest of the topics used in the prompt were added manually from reading random samples. The prompt in this stage is shown below

'*The text is from an online discussion platform mostly posted by tenants. Classify the text [summary/text] with weight (summing to 1) into at most three and at least one of the categories in utility issues or pest issues or mold issues or interior decoration issues or fee dispute issues or noise complaint issues or evicted by landlord issues or rent increase issues or deposit dispute issues or pet issues or landlord harassment issues or personal income decrease issues or sublease issues or covid risk issues. Return the results in the format of topic: weight separated by semicolon. order returned topics from highest to lowest weight.*'

Then, we prompted GPT to give the top three topics out of 15 topics that are the most pertinent to the text and at the same time asked it to provide weights to each topic that sum up to 1 for each text.

Regarding the weights, although GPT is asked to return the weight of the top three topics, such as "{utility: 0.7; evicted by landlord: 0.2; deposit dispute: 0.1}", we did NOT use the exact numerical values but rather the order of the topics. The purpose of asking GPT to return the weights is to ensure that the order and weights are consistent and that the returned order is accurate. Meanwhile, it's important to note that the exact mechanism by which GPT determines the weight during the process remains unclear.

When prompting with summary, the summary obtained from the first prompt for each text was inputted into the second prompt at the square bracket position. In the without summary case, which is the method we adopt for the results, the second prompt is the single prompt used and the original text is inputted into the square bracket.

The instruction in the prompts was aimed to be as specific as possible, including specifications on how weights should be assigned, the most and least number of topics allowed for each classification, and the output format. This effort yields standardized output for pipeline analyses. The topic names input into this prompt were also aimed to be as specific as possible to allow GPT to internally extract an unambiguous embedding for the best classification (Mu et al., 2023). Similar topics were then combined in all later analyses for the readability of the results, which gave 12 topics in total.

Moreover, when querying the LLM's, we adopt the default parameters[3]. The rationale for adopting a parameter-tuning free method in this study is such that this framework could be easily adopted by the general social science community, without requiring much computer science expertise. One important parameter is that temperature is set by default to 1. Temperature is a parameter that controls the randomness or creativity of the LLM's response. While the result of this study is based on response when temperature is set to a high value (1.0), we also conduct further experiments on a lower value of temperature (0.2) to verify the suitability of using temperature = 1 in this study and discuss the observations in the "results" section.

*Validation*

Following the framework outlined in the previous section, each text was modeled as a weighted combination of three topics and classified as the main topic. The validation set consists of 206 random samples, with around 10 from each topic. The samples were manually labeled with one single topic that's most representative of the post by two reviewers with backgrounds in both computer science and social science, specifically the housing topic. The first round of independent labels resulted in the Fleiss Kappa score of 0.72. As there are 206 samples, all the disagreements between the two reviewers in labeling were reconciled by discussions.

We measured the overall accuracy by the proportion of texts that were correctly classified by one of the topics assigned by GPT. We also broke down the accuracy into recall @ 1, recall @ 2, and recall @ 3 (Jonathan et al, 2004). Recall @ k measures the proportion of posts which GPT correctly classifies within their top k output. For example, if the classification provided by GPT with the highest weight (i.e. first place) coincided with the label of a text, that would imply a very accurate classification. The percentage of texts where this happens was defined as Recall @ 1, which was the proportion of texts whose classification by GPT with the highest weight was correct according to the validation label. Recall @ 2 and recall @ 3 were computed with the same logic but focusing on the classification by GPT with the second place and third place in the return. For instance, if the GPT's response is {utility, evicted by landlord, deposit dispute}, and the true label is deposit dispute, recall @3 will count the return as 1, while recall @1 and 2 will calculate it as 0. Similar applications about recall score can also be found in recommendation systems. (Li et al., 2021)

We also quantified the level of consistency of classifications made by GPT over the validation set by computing the Fleiss' kappa (Fleiss, 1971). Fleiss' kappa is a statistical measure of the level of agreement between multiple raters when making classifications. We ran GPT three times on the validation set and treated each trial as an individual rater when computing the Fleiss' kappa.

---

[3] OpenAI API Reference Create chat completion https://platform.openai.com/docs/api-reference/chat/create

Recall and Fleiss Kappa capture two distinct aspects of GPT's performance. The recalls are computed to quantify the accuracy of GPT's classifications, and Fleiss' kappa is computed over the output of GPT ran three times to measure how replicable the classifications are by GPT over multiple runs.

Furthermore, we also aim to understand the characteristics of misclassifications made by GPT. To do this, we calculate the recall @ 1, 2, and 3 for each true topic in the validation set. This helps us assess the variability in GPT's ability to correctly classify different topics. Additionally, we qualitatively examine several misclassified posts made by GPT. This provides some examples and hypotheses in the discussion section.

## Results

### *Topic Modeling*

The topic model derived 11 different topics. Based on inferences on the most frequent terms, we attributed a topic name to each of the 11 topics shown in Table 1. The top three referred to personal circumstances (16.5%), neighbor-related issues (13.1%), and structural or health hazards such as mold and water leaks (12.7%). Other dominant topics referred to rental contracts, eviction, and pet issues.

| Topic Name | Most Frequent Terms |
|---|---|
| Utility | 'heat', 'water', 'gas', 'window', 'parking' |
| Landlord Harassment | 'Notice', 'management', 'tenant' |
| Mold and Health Hazards | 'water', 'leak', 'maintenance', 'mold' |
| Interior Decoration | 'wall', 'paint', 'carpet', 'place' |
| Fee Disputes | 'deposit', 'charges', 'fee', 'check', 'money' |
| Personal Circumstances | 'ask', 'go', 'look' |
| Noise Complaint | 'noise', 'night', 'week', 'hear' |
| Eviction | 'eviction', 'court', 'notice', 'covid' |
| Legal Advice | 'sign', 'end', 'july', 'renew', 'roommate', 'year' |
| Neighbors | 'neighbor', 'go', 'leave' |
| Pet | 'dog', 'pet', 'cat', 'animal', 'inspection' |

Table 1. Topics from the LDA model and high-frequency words

### *Model Selection and Prompt Engineering*

Figure 2 compares the recalls @ 1, 2, 3 of different models and when adding a summary step or not. From Figure 2, we find GPT-4 to outperform GPT-3.5 when both types of prompts are used. This result is aligned with OpenAI's report, which demonstrates GPT-4's superior performance compared to GPT-3.5 in a set of other tasks including completing the SAT exam (Achiam et al., 2023). We also discovered from the results that both GPT-3.5 and GPT-4 perform better on the validation set when no intermediate step of summarization is performed. This is the reflection of the powerful ability of both models to comprehend long and complicated texts, and as a result classification performance benefits from the preservation of all information when no summarization is performed.

Moreover, note that the recall @ 1 of GPT-3.5 and GPT-4 are very similar, and GPT-4's performance is much better when it comes to recalls @ 2 and 3. Further investigating the output of GPT-3.5, we notice that about half of the time, the model only outputs one topic as the classification even though we prompt it to output three topics that best describe the post, which caused the decrease in overall accuracy while recall @1 is not much decreased. This suggests that the superiority of GPT-4's performance comes into play when it is given the opportunity to provide more than one output for the task of post text classification.

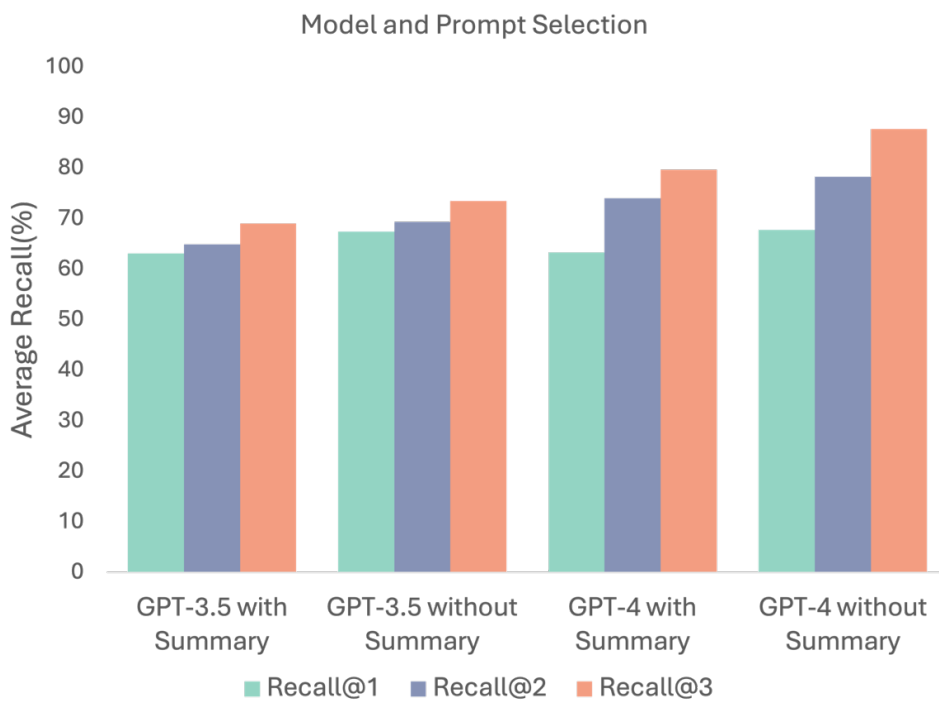

Figure 2. Recalls of Each Model Prompt Combination

Based on the experiment results in Figure 2, we deploy GPT-4 prompted without the summarization step. Using the output, we represented each post as a distribution of the 3 most pertinent topics. We finally assigned to each post a main topic, whose weight was the largest. Combining the assigned topic to each post with the geographic information that we have obtained before from entity recognition, we analyzed the geographic distribution of the topics. We also analyzed the temporal trends of the topics from 2015 to 2022 to explore dominant patterns in landlord-tenant relationships in different locations both before and during the pandemic.

*GPT-4 Classification*

Performance Validation GPT-4's classification yielded an overall accuracy (i.e. recall @ 3) of 86.70%. Recall @ 1 was 69.14%, recall @ 2 was 79.26%, and at 3 was 86.70%. After combining certain sub-topics into the same topic, Recall @ 1 was 70.76%. recall @ 2 was 79.79%, and at 3 was 87.23%. This indicates that GPT-4 is not predisposed to misclassifying to a similar sub-topic that belongs to the same overarching topic. The recalls for each individual topic labeled in the validation set are shown in Figure 3.

*Consistency Test*
The Fleiss' Kappa computed for GPT-4 over the validation set was 0.91. This value is very close to 1, which can be interpreted as falling in the range of almost perfect agreement between the raters. This result suggests that GPT-4 exhibits a high level of consistency when classifying post topics. This alleviates possible concerns on performance inconsistency based on the generative nature of GPT-4.

*Analysis of Model Temperature*
We also conduct the same classification task on the same validation set three times using GPT4 without summary step but setting the temperature to 0.2. The Fleiss' Kappa increased to 0.92 from 0.91. The overall average accuracy is 86.27%, with average recalls at 1,2,3 being 68.45%, 78.61%, 86.27% respectively. Note that recalls are similar to GPT-4 when temperature is default at 1. The above observations suggest that for the specific task in this study, using different temperature values has very minimal influence on both classification accuracy and consistency between multiple runs.

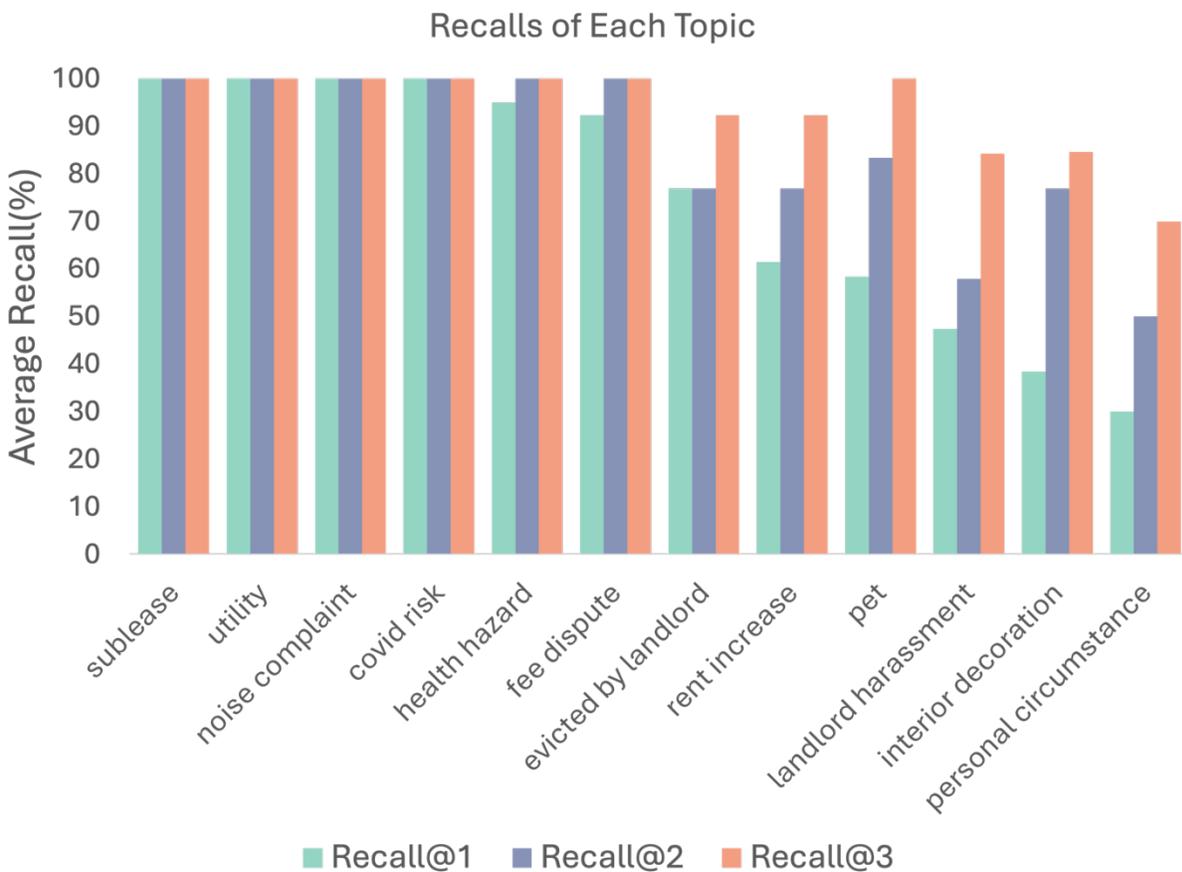

Figure 3. Recalls of Each Topic

*Geographic Analysis*

We found that 38.9% (N=1,676) of all posts have a U.S. state embedded in the text. For analysis in detail, we focused on two republican and two democratic states with the top number of posts: New York, California, Florida, and Texas. The distribution of topics within each of the four states is visualized in Figure 4. When combining geographic information, utility and deposit disputes were consistently the most prevalent topics in all four states. Evicted by Landlord issues were also considerably dominant in all states, especially in Florida. Landlord harassment issues are prevalent in all four states. The results imply that these issues remained to be the main pain points for tenants during the rental process throughout the US.

In terms of disparities among the four states analyzed, rent increase was a topic predominantly discussed in California but mentioned less in the other states. Fee disputes are particularly prevalent in California and Texas, while they are relatively uncommon in New York and Florida. Yet, sublease issues are commonly discussed in New York and not quite so in the rest of the states.

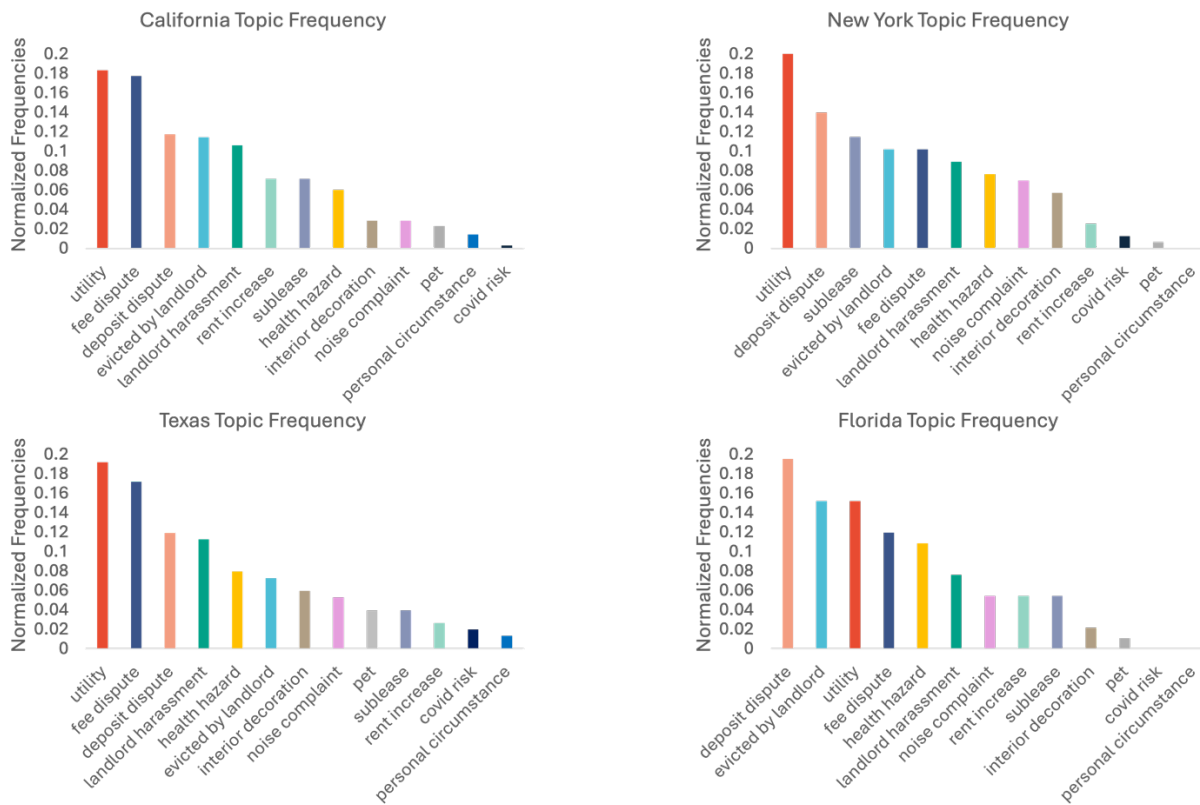

Figure 4. Topic Frequencies in Four States

Meanwhile, California does particularly well in having a low amount of discussion on health hazards while this remains a prominently discussed issue in Florida. On the other hand, California also has a high proportion of posts on rent increase issues, which is quite rarely discussed in the other three states. Surprisingly, texts that directly focused on COVID risks were very rare in all states.

*Temporal Analysis*

We begin analyzing the temporal trends of the number of posts from 2015 to 2023. As seen in Figure 5, the number of posts rocket-rises along with the onset of the COVID-19 pandemic and continues to surge as the Eviction Moratorium ends. This signifies that a greater amount of tenant concerns were caused by the pandemic and by the end of the Eviction Moratorium. To unpack the specific concerns that were the most correlated with the two events, we break down post frequencies by topic and analyze their temporal trends.

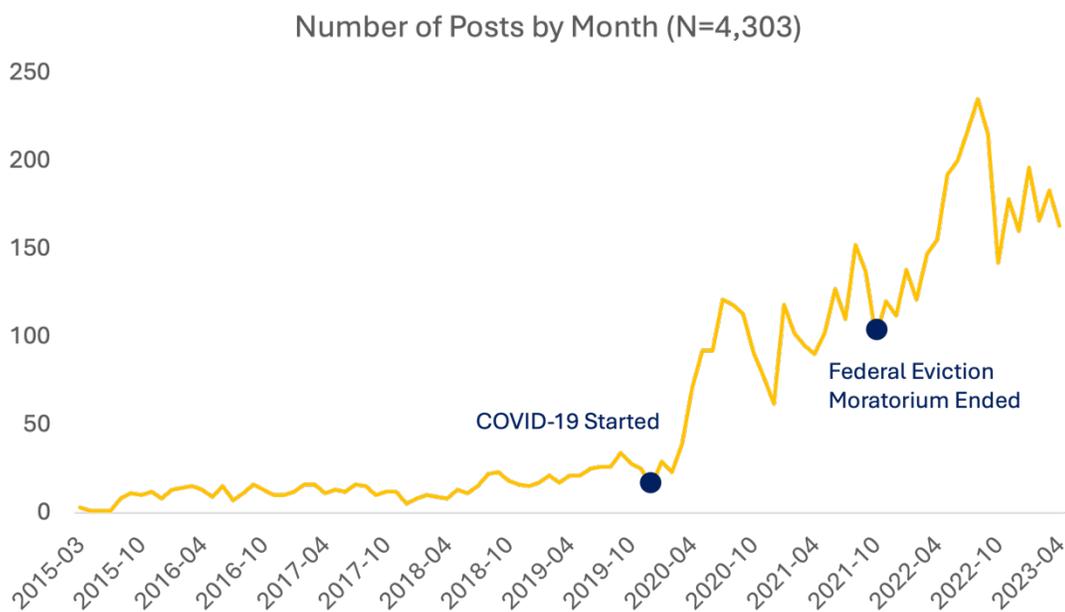

Figure 5. Temporal Trends in Post Frequency(N=4,303)

Regarding temporal trends of tenant concern topics, the frequencies of the topics saw a drastic increase in frequency from 2019 to 2020 when the pandemic began, indicating a conspicuous increase in tenant concerns correlated to the onset of the pandemic. The number of posts per unique user is about 1 throughout the time range in the study, showing that the increase in the number of posts was representative of an expanded scope of tenant concerns as opposed to an increased intensity of concerns by the same group of tenants.

Figure 6 shows the temporal trends of eight topics that are the most informative. Most topics' frequencies reached the zenith in 2021-2022 when the negative impact of COVID lockdowns and COVID-related health issues was the most severe. The upsurge starting from 2019 clearly manifests the increased anxiety and concerns that the pandemic has brought to tenants. Yet, the frequency of COVID risk issues remains quite low as compared to other topics. This portrays that COVID is more prone to causing concerns for tenants due to its social and economic repercussions but tenants are less concerned about the direct health consequences of COVID risks during the rental process (Tsai et al., 2022).

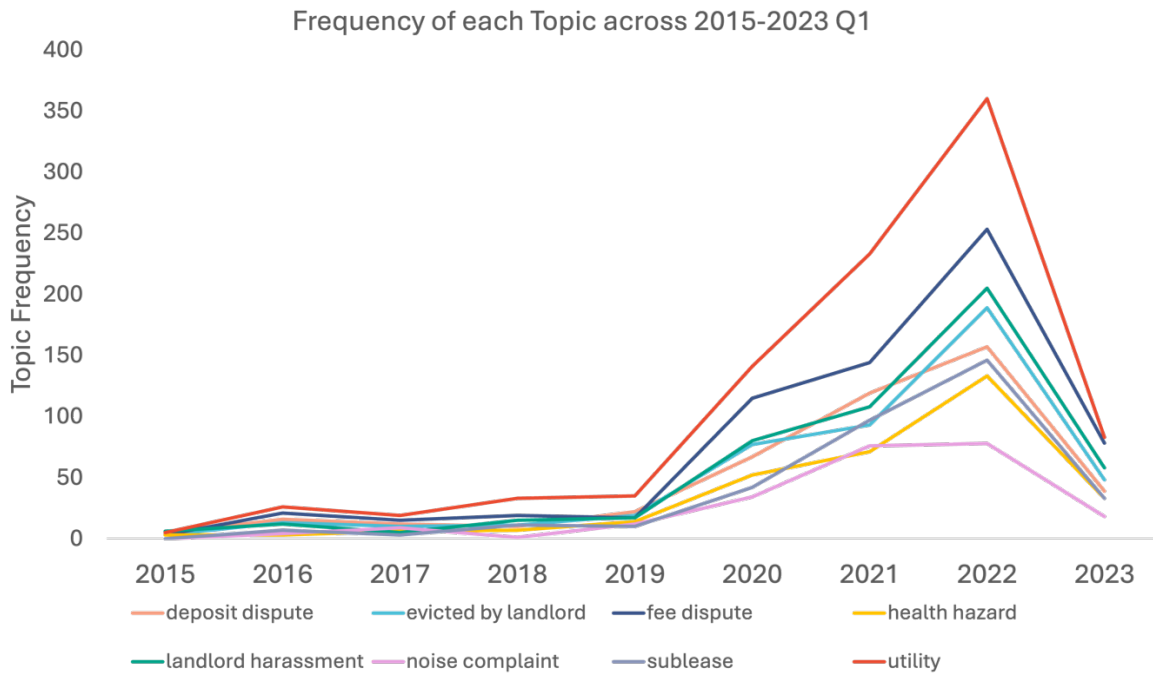

Figure 6. Temporal Trends in Topic Frequency (Note: Due to data accessibility, only the first quarter of 2023 (January to April) is included in the analysis)

We can see from the figure that utility issues and fee disputes suddenly stood out from the rest of the topics after 2019, becoming the most commonly mentioned issues in the posts. These issues should be given priority when aiming to address tenant concerns. Interestingly, the number of posts about noise complaints rose dramatically from 2020 to 2021. It then plateaus in frequency after 2021. This could be associated with the lockdown and stay-at-home circumstances during the pandemic, which has exacerbated the issue of noise complaints. (Torresin et al., 2022) found that lower psychological well-being was associated with the presence of more people at home. It was also found in a study that people are experiencing greater dissatisfaction due to noise created by other people in the same dwelling during the Pandemic (Sentop Dumen and Saher, 2020). This issue would beset tenants especially because they usually share a dwelling unit with other co-tenants. As neighbors also stayed at home more, noises from neighbors also sabotaged residents' comfort during the pandemic (Torresin et al., 2022). This hypothesis could also apply to the trends in health hazards and utility issues, which became more salient as tenants spent more time at home and when maintenance services were inhibited by the risk of infection.

Despite the Eviction Moratorium from Sept 2020 to July 2021, concerns regarding eviction represented by the posts continued to rise from 2020-2022. This trend exhibits the difficulty of implementing the new law, which was also expressed by Michael Trujillo, attorney of the Law Foundation of Silicon Valley, who claims that "It's going to be a real nightmare to try to educate tenants on what to do under the new California law (Sumagaysay, 2020)."

**Discussion**

The dataset used in this study demonstrates its strengths in various aspects. The dataset contains sufficient sample size to conduct meaningful analysis and the overall topics of discussion in the texts are all concentrated under the broader topic of tenant concerns. It is also geographically and temporally diverse, allowing us to draw insights from the texts from multiple perspectives.

The methodology of the study is also innovative in many aspects, including using a combination of topic modeling and zero-shot classification. Topic modeling, specifically LDA, enables exploratory analysis and informs topics present in the dataset at a low cost. The information obtained from the LDA model is then passed into GPT-4 for more accurate classification. The classification process by GPT-4 is completed by one single prompt, which reflects GPT-4's strong ability to comprehend complex and long texts.

A major limitation of this study, which is inherent from the generative nature of generative models and GPT specifically, is that the classifications are not replicable even when using the exact same prompts (Yan et al., 2023). Researchers hoping to replicate the results of this study might not be able to obtain the exact same results even with low temperatures. Researchers wishing to adopt the same methodology in this study would also need to be prepared for slight differences in the results each time the method is applied. Nevertheless, analyzing model consistency through Fleiss' Kappa shows that concerns about performance consistency have limited influence on the replicability of our framework.

We have sought to understand the potential causes of misclassification to better grasp the capabilities of the GPT model we utilized. This understanding will help us to avoid similar pitfalls in the future. In addition to requesting that GPT assign weights to each topic for consistency check, we've also prompted the model to "elucidate" why it selects a particular topic, aiming to better understand the mechanism behind GPT's classification. However, we must remain aware that GPT does not perform true reasoning—in other words, it does not understand the real meaning of text in the same way humans do. Therefore, any explanations it provides should not be considered evidence of its internal "reasoning" for a given classification. The purpose of applying this approach is to gain or test insights into the model's decision-making process even if the model may not understand the meaning of the text. According to our observations, in most circumstances, the explanation and logic of the choice appear coherent.

We specifically investigated topics where the recalls are lower than others, which include personal circumstances, interior decoration, and landlord harassment. We identify that posts predominantly on these topics usually involve the discussion of other topics, which they may get misclassified in. For instance, a person's concern is about not being added to the lease by her husband because her husband is moving to the new apartment early for some life changes, which we labeled as a post on personal circumstances. Yet, since the topic also mentions lease signing, GPT misclassified it as "fee dispute". Moreover, we find that of the posts misclassified in these topics are usually misclassified to topics of a similar nature. For instance, interior decoration is often misclassified to utility issues, and landlord harassment is commonly misclassified as evicted by landlord issues. Our observation implies further challenges for the GPT model to precisely differentiate posts of similar topics.

Moreover, fully comprehending the logic and drawing insightful conclusions remain a challenge for the model, leading to misclassifications (Wang and Demszky, 2023). Being a generative model, some of these misclassifications also mirror the model's generative characteristics and the biases inherent in its training data. This feature was shown after asking the model to provide some explanations. For example, a tenant complaint "Our large landlord company wants to change our physical locks to smart tech, and we prefer the security of physical keys," should be categorized as "landlord harassment". However, the model misclassified it as a "noise complaint", explaining that the "resident feels more secure with physical keys, suggesting the noise associated with the new smart tech system could be discomforting." Notably, there was no mention of noise in the original text. Such inherent flaws become evident when applying this model to our tasks. Whether these issues will be addressed in subsequent versions remains to be seen.

Another drawback of the method is its lack of systematic guidelines for prompt engineering, which is a limitation also shared by most generative LLMs. While following common best practices in literature may help enhance the process of prompt engineering, there are no systematic guidelines to do so. Trial-and-error of prompts might also be required to achieve a locally reasonable output performance (Hong and Davison, 2010). However, the specific procedure of trial-and-error and the definition of "reasonable" are vaguely defined in the literature, leaving many open decisions to be made by researchers hoping to adopt the method in this study. In our own design, we've come to realize that there's always room for improvement. For instance, the prompt "Summarize the following text in 10 words" could be too broad, potentially missing crucial information about the post. A better design could involve adding context to guide the model towards relevant aspects of tenant-landlord relationships.

A future direction for this study lies at the intersection of data science and Human-Computer Interaction (HCI) to bring together different areas of expertise. Hess and Chasins (2022) pioneered the collaboration between social and computer scientists through the use of a programming-by-demonstration tool for web automation, informing rental housing policy in the US. If we integrate the HCI perspective into similar research, we could explore how to incorporate feedback from housing research experts into AI systems to enhance their capabilities in specific tasks, potentially yielding more accurate outputs.

Another important future direction lies in the analysis of other sources of data to extract more generalizable results. One limitation of this study is that the current datasets are with selection bias because they are primarily from an online forum in English and within the US. Thus, it would be valuable to obtain data from other countries to compare tenant-landlord issues globally. It would also be important to investigate other sources of data as online forums are not the only venue where tenants express concerns. Some possible alternative sources of data include eviction court records, data from non-profit housing organizations, etc. Furthermore, it would be valuable to conduct experiments to compare the performance of other LLMs, such as Llama and Claude. Additionally, it would be intriguing to explore the advantages and disadvantages of various LLMs for similar tasks, thereby providing researchers with valuable insights to aid their decision-making processes.

Moreover, it would be informative to explore the landlord side and model landlord concerns from online forums to comprehensively understand the tensions arising in tenant-landlord relationships from both perspectives.

**Conclusions**

This study presents an analysis of posts from r/Tenant, revealing multifaceted insights into tenant-landlord relationships and tenant concerns. We found topics of consistent dominance across different states while distinguishing issues special to certain states. We also discovered meaningful temporal trends in many topics of tenant concerns and interpreted them under the context of the pandemic and the Eviction Moratorium, evaluating the implications of these social events and policies on tenant experiences. The study exhibits the power of web automation, machine learning, and modern Large Language Models to enrich research on tenant-landlord relationships, potentially shedding light on ways to create and reform housing policies and housing programs.